\documentclass{desyproc}

\begin{document}
	\title{Measurement of $^{144}\rm{Pr}$ beta-spectrum with Si(Li) detectors for the purpose of determining the spectrum of electron antineutrinos.}
	
	\author{{\slshape A.V. Derbin$^1$, S.~V.~Bakhlanov$^1$, I.~S.~Drachnev$^1$, I.~P.~Filipov$^1$, L.~A.~Lukyanchenko$^2$, I.~N.~Machulin$^2$, V.~N.~Muratova$^1$, N.~V.~Pilipenko$^1$, D.~A.~Semenov$^1$, E.~V.~Unzhakov$^{1}$}\\[1ex]
		$^1$NRC KI Petersburg Nuclear Physics Institute, 188300, Gatchina, Russia\\
		$^2$National Research Centre Kurchatov Institute, 123182 Moscow, Russia}
	
	\contribID{Derbin\_Alexander}
	
	\confID{13889}  
	\desyproc{DESY-PROC-2017-XX}
	\acronym{Patras 2017} 
	\doi  
	
	\maketitle
	
	\begin{abstract}
			Here we present the specifications of the newly developed 
		beta-spectrometer based on thick full absorption Si(Li) detector.
		The spectrometer can be used for precision measurements of various 
		beta-spectra, namely for the beta-spectrum shape study of $^{144}$Pr, which 
		is considered to be the most promising anti-neutrino source for sterile 
		neutrino searches.
		\end{abstract}



\section{Introduction}

Precision measurements of beta-spectra have always been and are still playing an
important role in several fundamental physical problems, predominantly in
neutrino physics.
It was the continuous shape of beta-decay spectrum, that has driven W.~Pauli
towards the neutrino hypothesis.
The initial beta-spectrum measurements for the purpose of determining the
neutrino mass were carried out by G.~Hanna and B.~Pontecorvo \cite{Hanna1949}
via gas counter that registered electrons produced by decays of tritium added
into the detector volume.

Magnetic and electrostatic spectrometers possess the superior energy resolution,
but at the same time such devices appear to be very complex and large-scale
instalments.
Since the electron free path at $3$~MeV (which is, basically, the maximum
beta-transition energy for long-living isotopes) does not exceed $2$~g/cm$^3$,
solid state scintillation and ionization detectors were effectively employed for
detection of electrons \cite{Simpson1985,Derbin1993}.
The main drawback of solid state scintillators is their relatively poor energy
resolution, which stands at approximately $10\%$ at $1$~MeV.
In case of semiconductor detectors there is a significant probability of
back-scattering from the detector surface that depends on the detector material.
The most widespread silicon-based semiconductors have the backscattering
probability of the order of $~10\%$ for $100$~keV electrons at normal incidence
\cite{Derbin1997}.
In case of electron energies above $1$~MeV and high $Z$ detector materials, it
also becomes important to take the bremsstrahlung into account.

The considered Si(Li) spectrometer was developed for precision measurement of
$^{144}$Ce-$^{144}$Pr beta-spectrum in order to determine the antineutrino
energy spectrum.
The $^{144}$Ce-$^{144}$Pr antineutrino source will be used for experimental
sterile neutrino searches by Borexino SOX collaboration \cite{Bellini2013}.

\begin{figure}[ht!]
	\centerline{\includegraphics[width=0.45\textwidth, height=0.3\textheight]{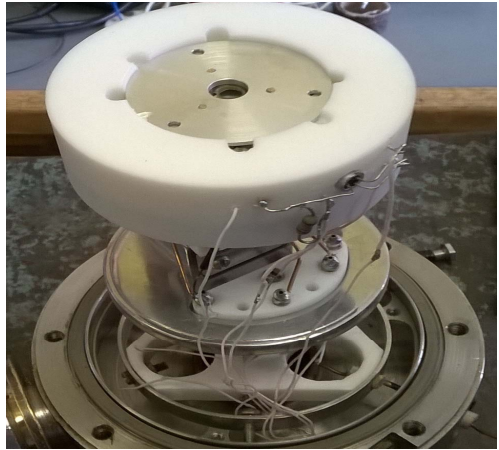}}
	\caption{The photo of the beta-spectrometer with the full-absorption Si(Li) detector.}
	\label{fig:1}
\end{figure}

\section{Experimental setup}

The photo of the experimental setup based on full absorption Si(Li) detector  is shown in Fig.\ref{fig:1}.
The sensitive region of the Si(Li) detector fabricated at NRC KI PNPI has a diameter about $15$~mm and a thickness of $6.5$~mm.
These dimensions ensure effective absorption of electrons with energies up to $3$~MeV.
The detector was equipped with an tungsten collimator with a diameter of $14$~mm and a thickness of $2$~mm.
The negative bias voltage of $1$~kV was applied directly to the gold coating of the detector.
The energy resolution determined for $59.6$~keV gamma-line of $^{241}$Am turned out to be $\text{FWHM} = 900$~eV.

The entire setup was placed inside the vacuum cryostat and cooled down to the liquid nitrogen temperature.
The detector was equipped with charge-sensitive preamplifier with resistive feedback and cooled field-effect transistor.
As noted above, the spectrometer was designed with intent of measuring the beta-spectrum of $^{144}$Pr in the energy range of $(0 - 3)$~MeV.
In order to perform measurements in such a broad dynamic range, the Si(Li) detector was equipped with two separate spectrometric channels, each with
spectrometric amplifier and 14-bit analogue-to-digital converter set up as a standalone module.
Channel settings were adjusted to register events within $(0.01 - 0.5)$~MeV and $(0.05 - 6.0)$~MeV energy intervals. 
The choice of the upper energy limit at $6$~MeV was conditioned by our intention to monitor the possible alpha-activity of the sample under investigation.  A dedicated DAQ control software allows one to acquire and record two 16000-channel spectra from the Si(Li) detector.

\section{Results}

\begin{figure}[ht!]
	\centerline{\includegraphics[width=0.45\textwidth, height=0.3\textheight]{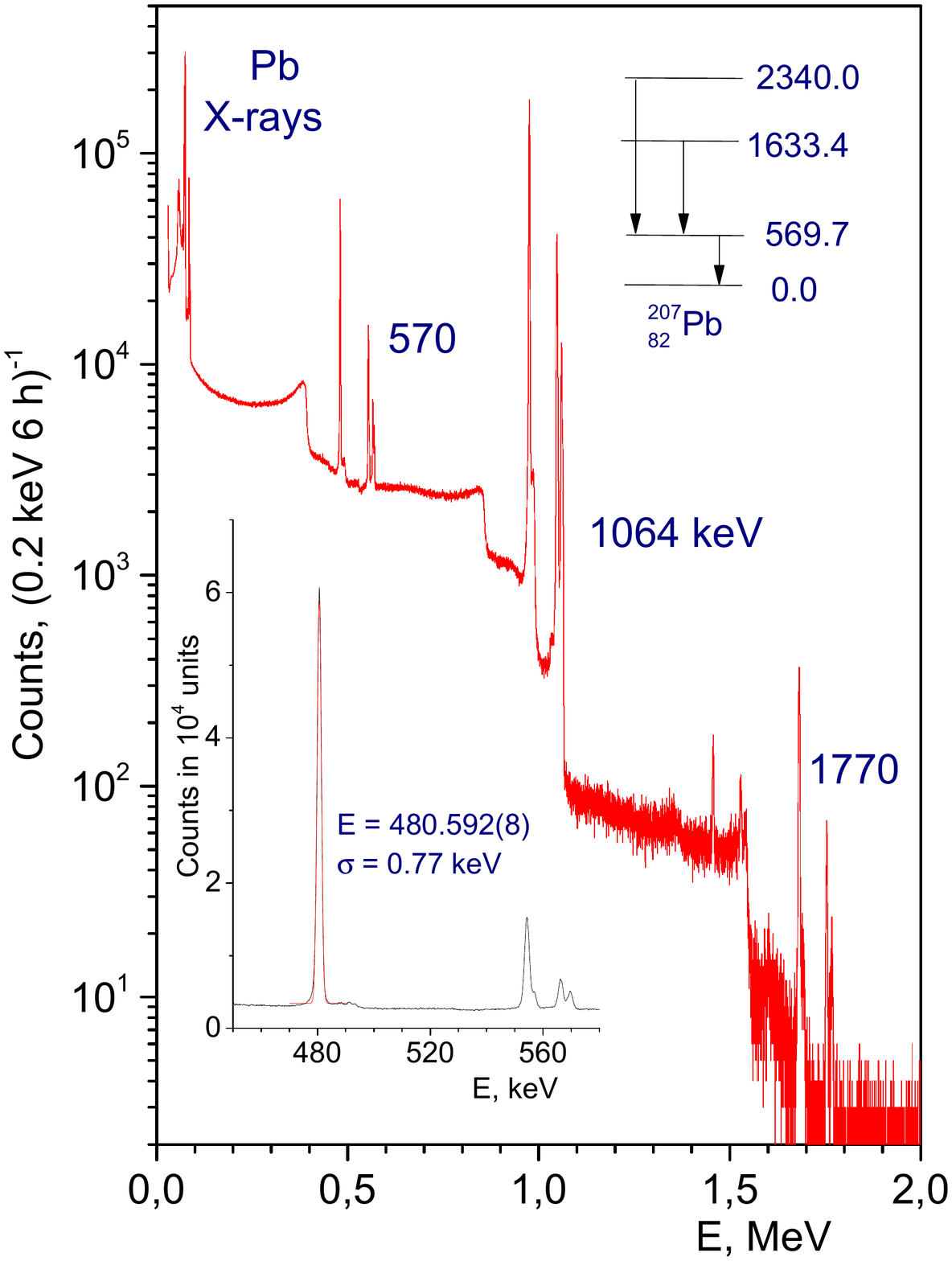}}
	\caption{The spectrum of $^{207}$Bi source measured with the Si(Li) detector in energy range of $(0.01 - 2.0)$~MeV.
		The inset shows the electron peaks corresponding to internal conversion from \mbox{$K$-}, \mbox{$L$-}, \mbox{$M$-}shells of the $570$~keV nuclear level.}\label{Fig:MV}
	\label{fig:2}
\end{figure}

In order to determine the main characteristics of the spectrometer we used a $^{207}$Bi source, providing gamma-rays, X-rays and conversion and Auger electrons.
The $^{207}$Bi spectrum, measured with the Si(Li) detector, is shown in Fig.\ref{fig:2} for the intervals $(0.01 - 2.0)$~MeV and $(450 - 580)$~keV, respectively.
The $^{207}$Bi source with an activity of $10^4$~Bq was placed inside the vacuum cryostat at a distance of $14$~mm from the Si(Li) detector surface.
Three of the most intense $^{207}$Bi gamma-lines have energies of $569.7$~keV, $1063.7$~keV and $1770.2$~keV and are emitted with probabilities of $0.977$, $0.745$ and $0.069$ per single $^{207}$Bi decay, respectively.
The corresponding peaks of conversion electrons form \mbox{$K$-}, \mbox{$L$-} and \mbox{$M$-shells} are clearly visible in the spectrum in Fig.\ref{fig:2}.
The electron energy resolution determined via $480$~keV line is $\text{FWHM} = 1.8$~keV.

\begin{figure}[ht!]
		\centerline{\includegraphics[width=0.45\textwidth, height=0.3\textheight]{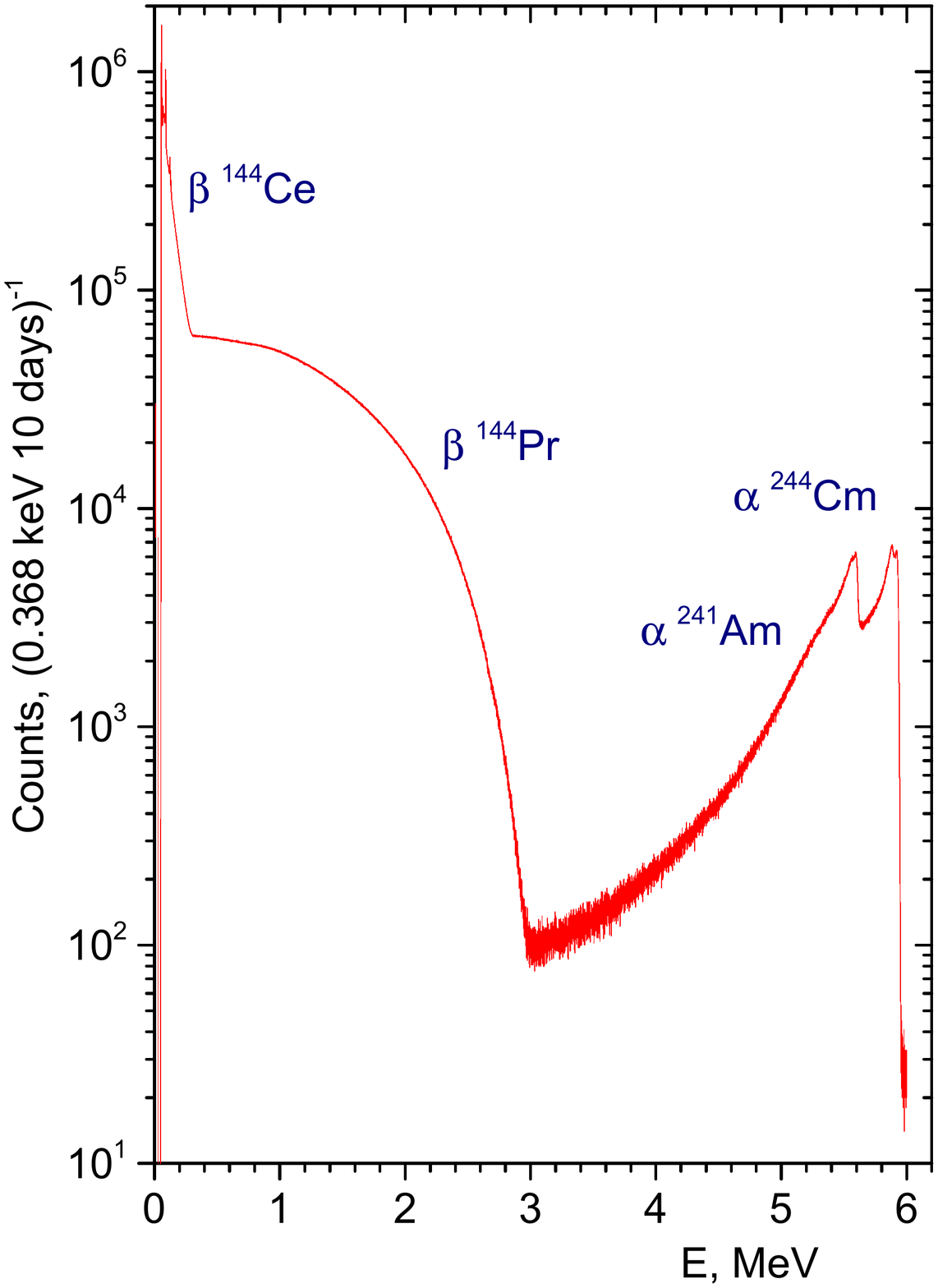}}
	\caption{The spectrum of $^{144}$Ce-$^{144}$Pr measured with the Si(Li) detector.}
	\label{fig:3}
\end{figure}

The low-energy part of the spectrum was used to evaluate the thickness of 
non-sensitive layer on the surface of Si(Li) detector. 
This area contains a set of peaks corresponding to Pb X-rays from \mbox{$K$}- 
and \mbox{$L$-series} and Auger electrons.
The observed position of $56.94$~keV Auger peak ($e_{KL_{1}L_{2}}$) appeared to be $56.22$~keV.
Inclusive of the golden coating thickness of ($500$~A$^\circ$), the measured $59$~keV electron energy loss of $720$~eV corresponds to $4700$~A$^\circ$ of non-sensitive layer.

The energy spectrum of the $^{144}$Ce-$^{144}$Pr source, measured with the Si(Li) detector during 10 days, is shown in Fig.\ref{fig:3}.
The total spectrum contains gamma- and electron peaks. The spectrum contains characteristic X-ray lines of Np at $13.9$~keV, $17.8$~keV and $20.8$~keV, produced by trace amounts of alpha-decaying $^{241}$Am contained in the $^{144}$Ce source. 
The peaks at $59.6$~keV, $80.1$~keV and $133.5$~keV are related to the gamma-transitions of $^{237}$Np and $^{144}$Pr nuclei. 
The most intense peaks at $35.8$~keV and $41$~keV correspond to \mbox{$K_{\alpha_1, \alpha_2}$-} and \mbox{$K_{\beta}$-lines} of Pr X-rays. The spectrum of electrons from beta-decay of $^{144}\text{Ce} - ^{144}\text{Pr}$ consists of also conversion and Auger electrons from 133.5~keV gamma-transition. 
After the main measurements were completed the thin silicon detector was mounted and $\gamma$- and X-rays activity  was measured by Si(Li)-detector in anticoincidence with thin Si-detector. 
Inclusion of this spectrum into the analysis allows us to account for overall gamma- and X-ray contribution.

The differential energy spectrum of electrons in $\beta-$decay is discribed as

\begin{equation}\label{beta-spectrum}
N(W)dW \sim pW(W-W_0)^2H(W)F(Z,W)L_0(Z,W)C_{A,V}(Z,W)S(Z,W)G(Z,W)B(W),
\end{equation}

where $W$ and $p$ are total energy and momentum of electron and $F(W,Z)$ is Fermi function, Additionally the following correction factors have to be taken into account: finite size of the nucleus correction for electromagnetic $L_0(Z,W)$ and weak $C_{A,V}(Z,W)$ interaction; screening corrections of the nuclear charge by electrons $S(Z,W)$; radiative corrections $G(Z,W)$ and weak magnetism correction $B(W)$. 
The antineutrino spectrum  can be calculated after the parameters of shape factor $H(W)$ will be extracted from the analysis of experimental spectrum. 

\section{Conclusion}

The beta-spectrometer with of 6.5 mm thick Si(Li)-detector has been developed.
The spectrometer can be used for precision measurements of the beta-spectrum shapes of various radioactive nuclei, in particular to measure the beta-spectra of $^{144}$Pr, which is the most promising antineutrino source for searching for neutrino oscillations to a sterile state.

\section{Acknowledgments}

This work was supported by the Russian Science Foundation (grant 17-12-01009) and in part by the Russian Foundation of Basic Research (grants 17-02-00305A, 16-29-13014ofi-m and 15-02-02117A)


\begin{footnotesize}

\end{footnotesize}


\end{document}